\documentclass[conference]{IEEEtran}

\ifCLASSINFOpdf
\usepackage[pdftex]{graphicx}
\else
\fi

\usepackage{url}
\usepackage[]{fancyhdr} %
\newcommand{\changefont}{\fontsize{9}{9}\selectfont}
\usepackage[cmex10]{amsmath}
\usepackage{multirow}
\usepackage{amssymb}
\usepackage[noadjust]{cite}
\usepackage[acronym]{glossaries}
\usepackage{nomencl}
\usepackage{upgreek}
\usepackage{nomencl}
\makenomenclature
\usepackage[dvipsnames]{xcolor}

\IEEEoverridecommandlockouts

\newacronym{DER}{DER}{Distributed Energy Resource}
\newacronym{DG}{DG}{Distributed Generator}
\newacronym{DGs}{DGs}{Distributed Generators}
\newacronym{MG}{MG}{Microgrid}
\newacronym{MMGs}{MMGs}{Multi-Microgrids}
\newacronym{MGs}{MGs}{Microgrids}
\newacronym{MILP}{MILP}{Mixed Integer Linear Programming}
\newacronym{RES}{RES}{Renewable Energy Source}
\newacronym{ESS}{ESS}{Energy Storage System}
\newacronym{WISE}{WISE}{Waterloo Institute for Sustainable Energy}
\newacronym{NPC}{NPC}{Net Present Cost}
\newacronym{OM}{OM}{Operation and Maintenance}
\newacronym{PV}{PV}{Photovoltaic}
\newacronym{GHI}{GHI}{Global Horizontal Irradiance}
\newacronym{NPV}{NPV}{Net Present Value}
\newacronym{RC}{RC}{Remote Communities}
\newacronym{SOC}{SOC}{State-of-Charge}
\newacronym{BAU}{BAU}{Business as Usual}
\newacronym{GHG}{GHG}{Greenhouse Gas}

\usepackage{etoolbox}
\renewcommand\nomgroup[1]{%
  \item[\bfseries
  \ifstrequal{#1}{I}{Subscripts and Sets}{%
  \ifstrequal{#1}{P}{Parameters}{%
  \ifstrequal{#1}{V}{Variables}{}}}%
]}

\nomenclature[I,01]{$I$}{Set of all generation and storage capacities $I=\{i\}$ }
\nomenclature[I,02]{$E$}{Subset of existing diesel generators $E=\{e\} \in I$ }
\nomenclature[I,03]{$N$}{Subset of new diesel generators $N=\{n\} \in I$}
\nomenclature[I,04]{$\Pi$}{Subset of RES and ESS $\Pi=\{p\} \in I$}
\nomenclature[I,05]{$S$}{Subset of solar panels $S=\{s\} \in \Pi$}
\nomenclature[I,06]{$W$}{Subset of wind turbines $W=\{w\} \in \Pi$}
\nomenclature[I,07]{$B$}{Subset of batteries $B=\{b\} \in \Pi$}
\nomenclature[I,08]{$F$}{Subset of fuel cells $F=\{f\} \in \Pi$}
\nomenclature[I,09]{$\Xi$}{Subset of electrolizers $\Xi=\{\xi\} \in \Pi$}
\nomenclature[I,10]{$Q$}{Subset of hydrogen tanks $Q=\{q\} \in \Pi$}
\nomenclature[I,11]{$h$}{hours}
\nomenclature[I,12]{$y$}{years}

\nomenclature[P]{$\alpha$}{Temperature coefficient of power for solar panels [pu$/^\circ$C]}
\nomenclature[P]{$\eta^{Ch}$}{Efficiency of battery charging [pu]}
\nomenclature[P]{$\eta^{Dch}$}{Efficiency of battery discharging [pu]}
\nomenclature[P]{$\eta_{f}$}{Efficiency of fuel cell [pu]}
\nomenclature[P]{$\eta_{\xi}$}{Efficiency of electrolizer [pu]}
\nomenclature[P]{$\gamma$}{Solar generation reserves coefficient [pu]}
\nomenclature[P]{$\rho$}{Wind generation reserves coefficient [pu]}
\nomenclature[P]{$\varphi$}{Derating factor of solar panels [pu]}
\nomenclature[P]{$\delta$}{Depth-Of-Discharge (DOD) of a battery [pu]}
\nomenclature[P]{$G_{stc}$}{Incident solar irradiance on solar panels at standard conditions [kW/m$^{2}$]}
\nomenclature[P]{$c$}{O\&M cost [\$/kWh]}
\nomenclature[P]{$\mathcal{H}$}{Total ammount of representative hours - 288 [h]}
\nomenclature[P]{$P^d$}{Power demand [kW]}
\nomenclature[P]{$G$}{Solar irradiance [kW/m$^{2}$]}
\nomenclature[P]{$t^{dch}$}{Time duration a battery can discharge continuously at a fixed power [h]}
\nomenclature[P]{$t^{ch}$}{Time duration a battery can charge continuously at a fixed power [h]}
\nomenclature[P]{$\tau$}{Solar cell temperature [$^\circ$C]}
\nomenclature[P]{$\tau_{stc}$}{Solar cell temperature at standard test conditions [$^\circ$C]}
\nomenclature[P]{$K$}{Unit cost of new diesel generators [\$/kW], RES [\$/kW] or battery [\$/kWh]}
\nomenclature[P]{$\mathcal{R}$}{Rated capacity of existing and new diesel generators [kW], RES [kW], or battery [kWh]}
\nomenclature[P]{$\mathcal{C}$}{Total number of cycles of charge and discharge of a battery}
\nomenclature[P]{$\beta$}{Demand reserves coefficient [pu]}
\nomenclature[P]{$\mathcal{A}$}{Total number of hours available in an average year for diesel generators [h]}
\nomenclature[P]{$\lambda$}{Total number of representative days in a month - 30}
\nomenclature[P]{$S_{h}$}{Wind speed}
\nomenclature[P]{$V$}{Higher Heating Value of hydrogen [kWh]}
\nomenclature[P]{$l_{C}$}{Hydrogen compressor load [pu]}
\nomenclature[P]{$\overline{\vartheta},\underline{\vartheta}$}{Upper/lower limits for hydrogen tank [pu]}
\nomenclature[P]{$\mathcal{D}$}{Cost of diesel [\$/l]}
\nomenclature[P]{$\psi$}{Minimum load operating level for existing and new diesel generators [pu]}
\nomenclature[P]{$\theta$}{Remaining/useful life of existing/new diesel generators [h]}
\nomenclature[P]{$\mathcal{M}$}{A very large number}

\nomenclature[V]{$\hat{P}$}{Capacity addition of RESs [kW] or batteries [kWh]}
\nomenclature[V]{$\mathcal{I}$}{Total installed capacity of RESs [kW] or batteries [kWh]}
\nomenclature[V]{$P$}{Power generated or consumed by $i \in \{E,F,N,S,W\} $ or $i \in \{\Xi\} $ [kW], respectively}
\nomenclature[V]{$P^{ch}$}{Battery charging power [kW]}
\nomenclature[V]{$P^{dch}$}{Battery discharging power [kW]}
\nomenclature[V]{$\mathcal{N}$}{Number of types of RESs or batteries considered}
\nomenclature[V]{$u^{ch}$, $u^{dch}$}{ON/OFF state of battery charging or discharging}
\nomenclature[V]{$SOC$}{State-of-charge of battery [kWh] or hydrogen tank [kgh]}
\nomenclature[V]{$u$}{ON/OFF state of diesel generators}
\nomenclature[V]{$Z$}{Total Net Present Cost (NPC) [\$]}
\nomenclature[V]{$\mathcal{F}$}{Fuel consumption [litre]}

\begin{document}

%
\title{Integration of Renewable Energy Sources for Low Emission Microgrids in Canadian Remote Communities}

\author{
\IEEEauthorblockN{Enrique Gabriel Vera, Claudio Canizares, and Mehrdad Pirnia}
\IEEEauthorblockA{University of Waterloo, Waterloo, ON, Canada.\\
evera@uwaterloo.ca, ccanizares@uwaterloo.ca, mpirnia@uwaterloo.ca}
}


%





\maketitle
\thispagestyle{fancy}
\pagestyle{fancy}

\IEEEpeerreviewmaketitle

%
\title{Integration of Renewable Energy Sources for Low Emission Microgrids in Canadian Remote Communities}

\author{
\IEEEauthorblockN{Enrique Gabriel Vera, Claudio Canizares, and Mehrdad Pirnia}
\IEEEauthorblockA{University of Waterloo, Waterloo, ON, Canada.\\
evera@uwaterloo.ca, ccanizares@uwaterloo.ca, mpirnia@uwaterloo.ca}
}


\maketitle


\begin{abstract}
In recent years, the electrification of Canadian Remote Communities (RCs) has received significant attention, as their current electric energy systems are not only expensive, but are also highly polluting due to the prevalence of diesel generators. In addition, RCs' inherent geographic characteristics impose a series of challenges that must be considered when planning their electricity supply. Thus, in this paper, an optimization model for the long-term planning of RC Microgrids (MGs) including Renewable Energy Sources (RESs) and Energy Storage Systems (ESSs) is proposed, with the objective of reducing costs and emissions. The proposed model considers lithium-ion batteries and hydrogen systems as part of ESSs technologies. The model is used to investigate the feasibility of integrating RESs and ESSs in an MG in Sanikiluaq, an RC in the Nunavut territory in Northern Canada. The results show that wind resources along with solar and storage technologies can play a key role in satisfying RC electricity demand, while significantly reducing costs and Greenhouse Gas Emissions (GHG). In addition, insights on sustainable and affordable policies for RC MGs are provided.    
\end{abstract} 
\printnomenclature
\section{Introduction}
\noindent Remote Communities' (\acrshortpl{RC}) unique features such as distant location, extreme weather conditions, energy consumption patterns, limited availability of energy sources, and absence of connection to the bulk power system have made supplying their electricity needs a challenging problem. Currently, the main source of electricity in \acrshortpl{RC} is diesel generators, and therefore, due to their significant Operations and Maintenance (O\&M), transportation, and fuel costs, delivering electricity to them has become economically and environmentally expensive \cite{[1],[2],[3],[4],[7],[19]}.

The deployment of clean Microgrids (\acrshortpl{MG}) has been recommended to satisfy \acrshort{RC} electricity needs, as MGs have the potential to provide cheaper, cleaner and more flexible and reliable electricity using a wide variety of \glspl{DER}, including \glspl{RES} and \glspl{ESS} \cite{[4],[5],[6]}. In addition, given the current state of development of hydrogen systems and considerable reduction in their capital costs, there is a potential for integration of electrolizers and fuel cells in \acrshort{RC} \acrshortpl{MG} \cite{[4],[5],[13],[17]}.

The authors of \cite{[1],[2],[3],[6],[9],[10],[11],[12]} propose models and techniques to design and plan \acrshortpl{MG} for \acrshortpl{RC} using \glspl{RES} and \glspl{ESS}, while highlighting their benefits and advantages. In all these references, planning approaches for small \acrshortpl{RC} with consideration of the communities electrification needs are proposed, with wind and/or solar generation being considered in the planning horizon. Most of them propose a multi-year planning optimization approach to examine the economic and environmental impacts of \acrshort{RES} integration in Canadian \acrshort{RC} \acrshortpl{MG}, demonstrating that \acrshort{RES} integration with \acrshort{ESS} and an appropriate diesel capacity can result in significant cost savings. However, none of these publications consider hydrogen storage systems as part of the \acrshort{ESS} technologies.

This paper proposes a long term planning model for \acrshort{RC} \acrshortpl{MG} with \glspl{RES} and \glspl{ESS}, including hydrogen systems. The proposed mathematical model investigates the feasibility of integrating such technologies in the planning of an \acrshort{MG} in Sanikiluaq, an \acrshort{RC} in Nunavut, which is part of the Canadian northern territories. The paper is based on \cite{[14]}, which is a non-per-reviewed technical report, with limited reach and validation. The proposed model includes a wide variety of renewable and nonrenewable generation resources and \glspl{ESS}, such as hydrogen systems and lithium-ion batteries, which makes it stand out from other approaches available in the literature. In addition, due to its linear characteristics, possible solutions can be evaluated in a fast, reliable, and inexpensive way to support energy planners in studying various planning alternatives. Finally, appropriate operating reserves are included to accommodate uncertainties associated with demand, solar, and wind generation. 

In order to asses the impact of different technologies, several planning scenarios with various combinations of resources are considered. The results of the long term planning for each scenario are compared in terms of economic, environmental, and other technical indices. The analysis includes an evaluation of the impact of \glspl{RES} and \glspl{ESS} in Canadian RCs, while quantifying the potential benefits of their implementation to support Canada's decarbonization goals.

The rest of the paper is organized as follow: In Section II, the optimization model proposed for the long term planning of \acrshort{RC} \acrshortpl{MG} is explained. In Section III, all the required data to apply the model in the community of Sanikiluaq are provided. Section IV presents and discusses the results of the long term planning model for the Sanikiluaq \acrshort{MG}. Finally, the main conclusions of the presented work are highlighted in Section V. 

\vspace{-0.2cm}
\section{Model Description}
\noindent The proposed planning model is formulated using an optimization framework to plan the energy resources in RCs using diesel, wind, and solar generators, in combination with battery and hydrogen \glspl{ESS}. In addition to planning constraints restricting the type and amount of generation in different years, the model contains operational constraints with binary variables associated with the hourly on/off status of diesel generators, and the charging and discharging status of batteries and hydrogen storage systems. Integer variables are used to prescribe the quantities of different technologies for economic evaluation, while the variables representing the generation in kW, \gls{SOC} of batteries in kWh, and hydrogen storage systems in kgh are continuous. The model can therefore be characterized as a \gls{MILP} problem as described in detail next.

In the equations that follow, all generators and storage capacities are part of the set $I=\{i\}$, while existing diesel generators, new diesel generators, and \gls{RES} and \gls{ESS} form the sub-sets $E=\{e\}$, $N=\{n\}$, and $\Pi=\{p\}$, respectively. The subset $\Pi$ includes subsets of solar panels $S=\{s\}$, wind turbines $W=\{w\}$, batteries $B=\{b\}$, fuel cells $F=\{f\}$, electrolizers $\Xi=\{\xi\}$, and hydrogen tanks $Q=\{q\}$. Finally, $y$ is the index used for years, and the index $h$ is used for representative hours.

\subsection{Objective Function}\label{times_expl}
\noindent The following objective function represents the summation of the \gls{NPC} of the capital, fuel, and O\&M costs of the generators in the \acrshort{MG}: 
\begin{equation}
\begin{aligned}
  Z =   
\sum_{i,y \hspace{1mm} \forall i\in \{N,\Pi\}}K_{i,y}\displaystyle \hat{P}_{i,y}+
\sum_{i,y,h \hspace{1mm} \forall i\in\{E,N\}}\lambda\mathcal{D} \mathcal{F}_{h,y,i}\\
+\sum_{i,y,h \hspace{1mm} \forall i\in\{E,F,N\}} \lambda c_{i}P_{i,y,h}+ 
\sum_{i,y \hspace{1mm} \forall i \in \Pi-F}\mathcal{H}c_{i}\mathcal{I}_{i,y}
\end{aligned}
    \label{OBJFUNC}
\end{equation} 

\noindent where $K_{i,y}$ is the \gls{NPC} of the capital cost of a generation unit $i$, installed in year $y$; $\hat{P}_{i,y}$ is the amount of installed capacity of $i$ in year $y$; $\mathcal{D}$ is the cost of diesel fuel; $\mathcal{F}_{i,y,h}$ is the hourly diesel fuel consumption\footnote{ $\mathcal{F}_{i,y,h}$ is computed using the fuel curves available in \cite{[1]}, which are nonlinear and thus piece-wise linearization is used for their representation.}; $c_{i}$ is the hourly O\&M cost; $P_{i,y,h}$ is the generated power from generator $i$, in year $y$ and hour $h$; and $\mathcal{I}_{i,y}$ is the total installed capacity of generator $i$ in year $y$. Note that the total capital cost in the first term of (\ref{OBJFUNC}) is defined over generators $\Pi$ and $N$, and the fuel cost is considered only for $N$ and $E$. Factors $\lambda=30$ and $\mathcal{H}=288$ are used the carry out the calculations over the whole year, where  $\mathcal{H}$ is total number of representative hours in a year, \textcolor{black}{i.e., $24$ (average hours/month) $\times$ $12$ (months) = $288$ hours}, representing a 24-hours day for each of the 12 months, and $\lambda$ indicates the representative number of days in a month. The units of each parameter and variable are discussed in Section \ref{SEC3}.
\subsection{Constraints}
\subsubsection{Installed Capacity}
The total installed capacity $\mathcal{I}_{i,y}$ for $i \in \{N,\Pi\}$ each year $y$ is calculated by updating the total installed capacity of the previous year $\mathcal{I}_{i,y-1}$, as follows:
\begin{equation}\label{eqInstCap}
    \mathcal{I}_{i,y}=\hat{P}_{i,y}+I_{i,y-1}
    \hspace{0.3cm} \forall i\in \{N,\Pi\} ,y
\end{equation}
\noindent where the capacity additions $\hat{P}_{i,y}$ for $i \in \{N,\Pi-S\}$ at each year $y$ is defined by the product of the number of generators added each year $\mathcal{N}_{i,y}$, and their respective individual rated capacity $\mathcal{R}_i$, as follows: 
\begin{equation}
   \hat{P}_{i,y}=\mathcal{N}_{i,y} \mathcal{R}_{i}
   \hspace{1cm} \forall i\in \{N,\Pi-S\} ,y
\end{equation}
\noindent  Note that $\mathcal{N}_{i,y}$ is an integer variable for $i \in \Pi-S$, and is a binary variable for $i \in N$, since only one diesel generator of predefined capacities can be added to the generation portfolio each year. Finally, the capacity additions of solar $\hat{P}_{s,y}$ is a continuous variable, as the installation of solar panels is more versatile, since power fractions can be accommodated in practice. 
\subsubsection{Supply-Demand Balance}
\noindent The summation of the power generated by existing and new diesel generators $P_{e,y,h}$ and $P_{n,y,h}$, solar panels $P_{s,y,h}$, wind turbines $P_{w,y,h}$, fuel cells $P_{f,y,h}$, and battery storage discharge $P_{b,y,h}^{dch}$ should satisfy the total consumers' demand $P_{y,h}^{d}$, the battery storage charge $P_{b,y,h}^{ch}$, and the power consumed by the electrolizer $P_{\xi,y,h}$, at each hour $h$ and year $y$, as follows: 
\begin{equation}\label{eqSuppDem}
    \begin{split}
        \sum_{i \in \{E,F,N,S,W\}} P_{i,y,h}+\sum_{B}P^{dch}_{b,y,h}=P^{d}_{y,h}+\sum_{B}P^{ch}_{b,y,h}\\+\sum_{\Xi}P_{\xi,y,h} \hspace{0.5cm} \forall h,y
    \end{split}
\end{equation}
\subsubsection{Operating Reserves}
\noindent To accommodate the uncertainties associated with demand, solar, and wind generation, the rated capacity of existing diesel generators $\mathcal{R}_{e}$ and total installed capacity of new diesel generators $\mathcal{I}_{n,y}$ and fuel cells $\mathcal{I}_{f,y}$, plus batteries storage power capacity per hour $SOC_{b,y,h}$, have to be greater than the hourly consumers demand $P^{d}_{y,h}$ by a given factor $\beta$, and solar and wind generation by given factors $\gamma$ and $\rho$, respectively, for every hour during the planning horizon, as follows:
\begin{equation}
\begin{split}
    \sum_{E}R_{e}+
    \sum_{i\in\{F,N\}}\mathcal{I}_{i,y}+
    \sum_{B}&SOC_{b,y,h} \ge (1+\beta)P^{d}_{y,h} \\
   +\gamma\sum_{S}P_{s,y,h}
   &+\rho\sum_{W}P_{w,y,h}  \hspace{1cm} \forall h,y 
\end{split}
\end{equation}
\subsubsection{Diesel Generator Limits}
\noindent At every hour during the planning horizon, the power generated by diesel generators $P_{i,y,h}$ for $i \in \{E,N\}$ has to be less than or equal to the rated capacity of existing generators $\mathcal{R}_{e}$ and the total installed capacity of new diesel generators $\mathcal{I}_{n,y}$, and should also be greater than the minimum load operating level $\psi_{i}$ for $i \in \{E,N\}$, which is a factor of the rated capacity, as follows:
\begin{equation}
    P_{n,y,h} \leq \mathcal{I}_{n,y}u_{n,y,h}
         \hspace{1cm} \forall n,h,y 
         \label{NDG1}
\end{equation}
\begin{equation}
    P_{e,y,h} \leq \mathcal{R}_{e}u_{e,y,h}
    \hspace{1.6cm} \forall e,h,y 
    \label{EDG1}
\end{equation}
\begin{equation}
     P_{n,y,h} \geq \psi_{n}\mathcal{I}_{n,y}u_{n,y,h}
     \hspace{1cm} \forall n,h,y 
     \label{NDG2}
\end{equation}
\begin{equation}
    P_{e,y,h}\geq \psi_{e}\mathcal{R}_{e}u_{e,y,h} \hspace{1.2cm} \forall e,h,y 
    \label{EDG2}
\end{equation}
\noindent where $u_{i,y,h}$ for $i \in \{E,N\}$ is a binary variable indicating the operating on/off state of each generator. Equations (\ref{NDG1}) and (\ref{NDG2}) are nonlinear, and thus a common linearization technique is applied, as per \cite{[16]}. 
\subsubsection{Diesel Generator Service Life}
\noindent The useful life of new diesel generators and the remaining life of existing diesel generators $\theta_{i}$ for $i \in \{E,N\}$, in hours, is taken into account by computing their total amount of operating states $u_{i,y,h}$ for $i \in \{E,N\}$ during the planning horizon as follows:
\begin{equation}\label{life1}
    \sum_{h,y}\lambda  u_{i,y,h}\leq \theta_{i} \hspace{2cm} \forall i\in \{E, N\}
\end{equation}
\noindent Note that the factor $\lambda=30$ is used to represent the life of the generators over a year. \textcolor{black}{Therefore, their use is optimized and they get retired when reaching their limits.}
\subsubsection{Diesel Generator Availability}
\noindent This constraint is used to reflect the maintenance of existing and new generators during the planning horizon. Thus, a percentage of the total number of the hours available $\mathcal{A}$ in an average year is assigned for this purpose, as follows:
\begin{equation}
    \sum_{h}u_{i,y,h} \leq \mathcal{H}(1-\mathcal{A})
    \hspace{1cm} \forall i\in \{E,N\},y 
\end{equation}
\subsubsection{Solar Power Generation}
\noindent The solar power generation output is computed as a direct function of the hourly incident irradiance $G_{h}$, hourly cell temperature $\tau_{h}$, and derating factor $\varphi$, which is a scaling factor to account for effects of dust, wire loses, and other deviations of the solar output from its ideal value, as follows:
\begin{equation}\label{solarEq}
P_{s,y,h} = \varphi I_{s,y}\left(\frac{\displaystyle G_{h}} {G_{stc}} \right)\left[1+\alpha(\tau_{h}-\tau_{stc}) \right]  
\hspace{0.2cm} \forall s,y,h
\end{equation}
\noindent where $stc$ stands for standard test conditions. 

\subsubsection{Wind Power Generation}
\noindent The wind power is computed as a function of the hourly wind speed $S_{h}$ as follows:
\begin{equation}
  P_{w,y,h}=W(\mathcal{I}_{w,y},S_{h})
    \hspace{1.8cm} \forall w,y,h \hspace{-0.6cm}  
\end{equation}
\noindent where the power generated by every wind turbine is computed using its turbine power curve $W(\cdot)$ and the wind speed $S_{h}$ at every time-step \cite{[15]}.

\subsubsection{Battery SOC and Limits}
\noindent The following constraints compute the \gls{SOC} of the batteries as a function of the batteries' charge $P_{b,y,h}^{ch}$ and discharge $P_{b,y,h}^{dch}$ for every hour of operation $h$, considering the charging $\eta^{ch}$ and discharging $\eta^{dch}$ efficiency rates:
\begin{equation}
\begin{split}
SOC_{b,y,h+1}-SOC_{b,y,h}=\eta^{ch} P_{b,y,h}^{ch}-\frac{P_{b,y,h}^{dch}}{\eta^{dch}} \\   \hspace{1cm} \forall b,y,h    
\end{split}
\end{equation}
\begin{equation}
\begin{split}
  SOC_{b,y+1,1}-SOC_{b,y,\mathcal{H}}=\eta^{ch} P_{b,y,\mathcal{H}}^{ch}-  \frac{P_{b,y,\mathcal{H}}^{dch}}{\eta^{dch}} \\ \hspace{1cm} \forall b,y,h 
\end{split}
\end{equation}
\noindent The \gls{SOC} of batteries is subject to the following constraints reflecting the minimum and maximum capacity of the batteries: 
\begin{equation} \label{SOC_lim1} 
SOC_{b,y,h}\leq \mathcal{I}_{b,y} 
\hspace{1.2cm} \forall b,y,h   \hspace{-2cm}
\end{equation}
\begin{equation} \label{SOC_lim2}
SOC_{b,y,h}\geq \delta \mathcal{I}_{b,y}
\hspace{1cm} \forall b,y,h  \hspace{-2cm}  
\end{equation}
\noindent where $\delta$ is a factor to indicate depth of discharge of the batteries. The following constraints reflect the maximum charging and discharging limits respectively, and are functions of the depth of discharge $\delta$, the total installed battery capacity $\mathcal{I}_{b,y}$, and the continuous time duration of charging $t^{ch}$ and discharging $t^{dch}$, which are battery parameters chosen to keep reasonable equipment costs, while having adequate energy resources in a day:
\begin{equation} \label{SOC_lim3}
P_{b,y,h}^{dch} \leq \left(\frac{1-\delta}{t^{dch}} \right)\mathcal{I}_{b,y}
\hspace{0.5cm} \forall b,y,h \hspace{-1.8cm}
\end{equation}
\begin{equation} \label{SOC_lim4}
P_{b,y,h}^{ch} \leq \left(\frac{1-\delta}{t^{ch}} \right)\mathcal{I}_{b,y}
\hspace{0.5cm} \forall b,y,h \hspace{-1.8cm}
\end{equation}
\noindent Furthermore, the following constraints guarantee minimum charging/discharging power at a given hour:
\begin{equation} 
P_{b,y,h}^{dch}\geq u_{b,y,h}^{dch}
\hspace{1.2cm} \forall b,y,h \hspace{-2.2cm}
\end{equation}
\begin{equation}
P_{b,y,h}^{ch}\geq u_{b,y,h}^{ch}
\hspace{1.2cm} \forall b,y,h \hspace{-2.2cm}
\end{equation}
\noindent where $u_{b,y,h}^{dch}$ and $u_{b,y,h}^{ch}$ are binary variables indicating the battery operating states.

In order to prevent charging and discharging occurring at the same time, the following equation is used:
\begin{equation} \label{Ch_DCH} 
P_{b,y,h}^{dch}P_{b,y,h}^{ch}=0 
\hspace{1cm} \forall b,y,h \hspace{-2.2cm}
\end{equation}

\noindent which is not linear and is therefore substituted  by the following set of equations:
 \begin{equation}
    \hspace{1cm}
    P_{b,y,h}^{dch} \le u_{b,y,h}^{dch}\mathcal{M}
    \hspace{1cm} \forall b,y,h \hspace{-1cm}
 \end{equation}
 \begin{equation}
    \hspace{1.2cm}
    P_{b,y,h}^{ch} \le u_{i,y,h}^{ch}\mathcal{M}
    \hspace{1cm} \forall b,y,h \hspace{-1cm}
 \end{equation}
 \begin{equation}
    \hspace{1cm}
    u_{b,y,h}^{dch}+u_{i,y,h}^{ch} \le 1
    \hspace{1cm} \forall b,y,h \hspace{-1cm}
 \end{equation}
 \noindent where $\mathcal{M}$ is a very large number. Finally, the following constraint defines the life span of each battery for $\mathcal{C}$ cycles of charge and discharge:
\begin{equation}
    \sum_{h,y}(P_{b,y,h}^{ch}+P_{b,y,h}^{dch}) \leq \mathcal{C} \sum_{y} \hat{P}_{b,y}\\
    \hspace{0.6cm} \forall b,y,h   \hspace{-0.3cm}
\end{equation} 

\subsubsection{Hydrogen System}
\begin{figure}[t]
    \centering
    \includegraphics[width=\linewidth]{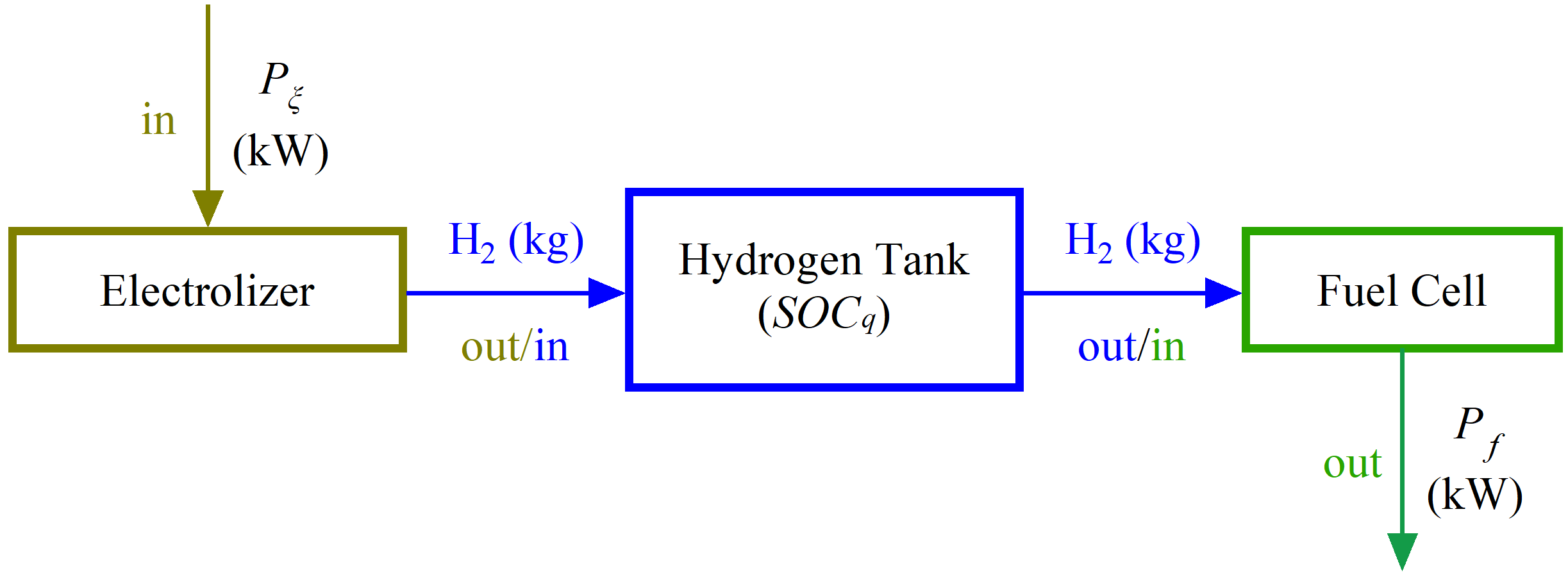}
    \caption{Schematic representation of a hydrogen storage system.}
    \label{Hydrogen}
\end{figure}

The hydrogen system is composed of an electrolizer, consuming electricity $P_{\xi,y,h}$ for generating the hydrogen that is stored at high pressure in tanks,  which is used later by the fuel cells to generate electricity $P_{f,y,h}$. A schematic representation of this process is presented in Fig. \ref{Hydrogen}. For this system, the \acrshort{SOC} of the hydrogen tank for every hour of operation $h$, $SOC_{q,y,h}$, is a function of the power generated by the fuel cells $P_{f,y,h}$ and the power consumed by the electrolizer $P_{\xi,y,h}$, which can be transformed into hydrogen consumption as follows \cite{[13]}:
\begin{equation}
\begin{split}
        SOC_{q,y,h+1}-SOC_{q,y,h}=\frac{1}{1+l_{C}}\frac{P_{\xi,y,h}\eta_{\xi}}{V}        -\frac{P_{f,y,h}}{V\eta_{f}}\\ \forall q,y,h    
\end{split}
\end{equation}
\begin{equation}
\begin{split}
        SOC_{q,y+1,1}-SOC_{q,y,\mathcal{H}}=\frac{1}{1+l_{C}}\frac{P_{\xi,y,\mathcal{H}}\eta_{\xi}}{V}-\frac{P_{f,y,\mathcal{H}}}{V\eta_{f}}\\ \forall q,y,h    
\end{split}
\end{equation}
where, for every year $y$, the hourly \acrshort{SOC} limits of the hydrogen tank are as follows:
\begin{equation}
         SOC_{q,y,h}\le \overline{\vartheta} \mathcal{I}_{q,y}
         \hspace{1cm} \forall q,y,h
\end{equation}
\begin{equation}
        SOC_{q,y,h} \ge \underline{\vartheta} \mathcal{I}_{q,y}
        \hspace{1cm} \forall q,y,h
\end{equation}
\noindent and $V$ is the Higher Heating Value of Hydrogen in kWh; $l_{C}$ is the hydrogen compressor load in pu; $\mathcal{I}_{i,y}$ is the net capacity of the hydrogen tank in kg; $\eta_{f}$ and $\eta_{e}$ are the efficiency of fuel cells and electrolizers, respectively; and $\overline{\vartheta}$ and $\underline{\vartheta}$ are per unit constants defining the maximum and minimum hydrogen tank limits.
In addition, the power generated by the fuel cells $P_{f,y,h}$ and the power consumed by the electrolizers $P_{\xi,y,h}$ need to be less than their total installed capacity $\mathcal{I}_{i,y}$ for $i \in \{F,\Xi\}$, as follows: 
\begin{equation}
    P_{i,y,h} \le \mathcal{I}_{i,y} 
    \hspace{1cm} \forall i\in\{F,\Xi\},y,h
\end{equation}

\vspace{-0.2cm}

\vspace{-0.2cm}
\section{Case Study}\label{SEC3}
The proposed model in Section II is used to investigate the feasibility of integrating RESs and ESSs in the planning of an MG in Sanikiluaq, an RC in Nunavut, which is part of the Canadian northern territories \cite{[15]}. The various parameters needed to apply the presented optimization model and their sources are provided next.
\vspace{-0.3cm}
\subsection{Electricity Demand}
\begin{figure}[t]
\centering
\includegraphics[width=\linewidth]{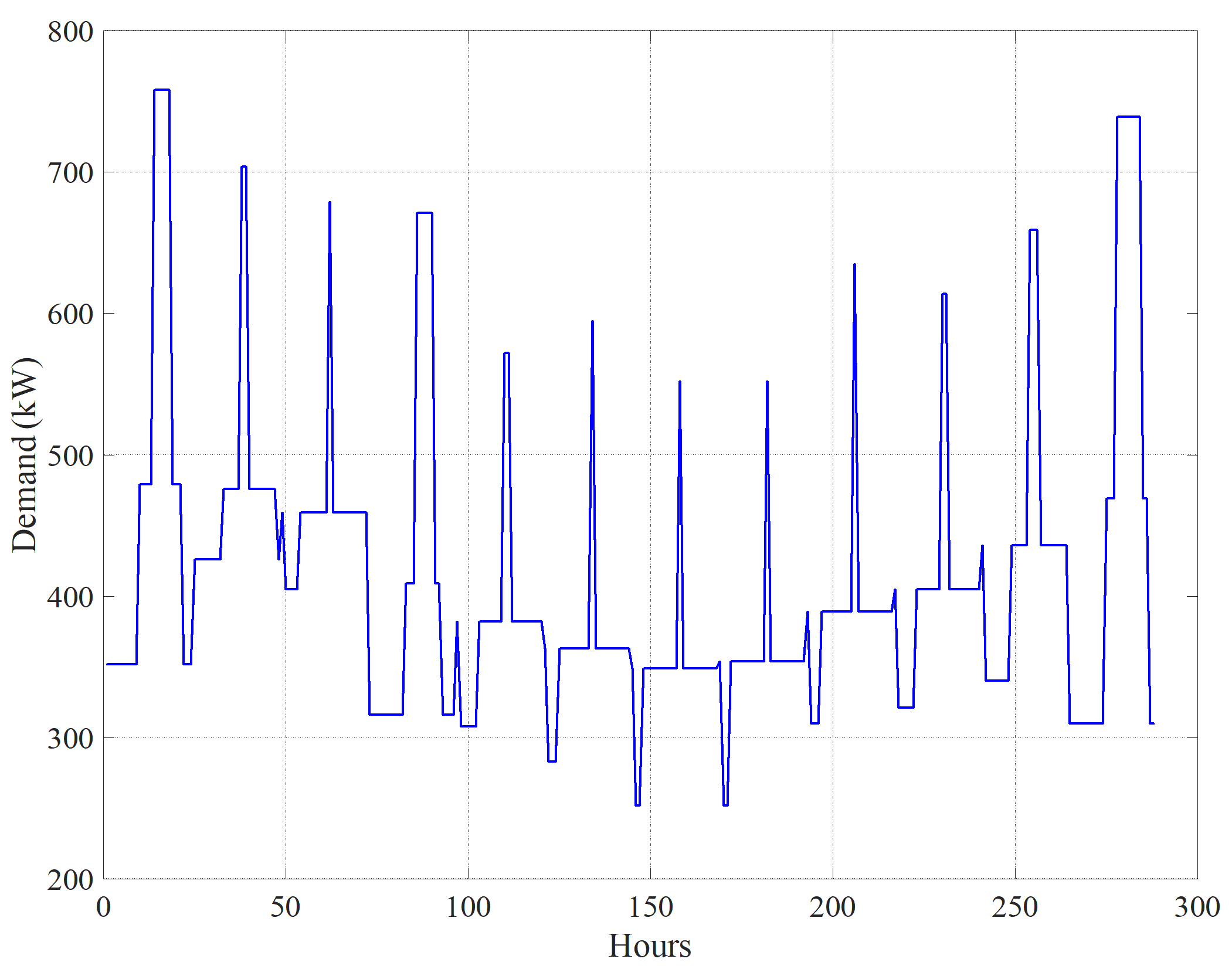}
\caption{Sanikiluaq's yearly average load profile \cite{[1],[15]}.}
\label{Load_day}
\end{figure}

The hourly load for the Sanikiluaq community was extracted from \cite{[1]} and \cite{[15]}. This data can be used to calculate the hourly averages for a year with 288 representative hours, as explained in Section \ref{times_expl}. The load is primarily residential and the corresponding demand profile is depicted in Fig. \ref{Load_day}.
\vspace{-0,2cm}
\subsection{Existing Diesel Generators}
The main characteristics of the existing diesel generators are presented in Table \ref{Data_Gen_San}. It is assumed that the minimum load of these generators is 40\% of their nominal power, i.e., $\psi_{e}=0.4$. In addition, generators 1, 2, and 7 are in stand-by mode during even years, whereas generators 3, 4, 5, and 6 are in stand-by mode during odd years, throughout the planing horizon. All generators, including those in stand-by mode, are assumed to act as reserves for the \acrshort{MG}, as per \cite{[1]} and \cite{[15]}. 
\begin{center}
    \begin{table}[t]
    \caption{Main generators' characteristics at Sanikiluaq \cite{[1],[15]}}
        \centering
        \begin{tabular}{|c|c|c|c|c|c|}
        \hline
        \multirow{2}{2em}{Gen.} & Capacity & Lifetime & a & b & c\\
        & [kW]  & [h] & [$l/h/kW^{2}$] & [$l/h/kW$] & [$l/h$]\\ 
        \hline
        1 & 330 & 35,339 & -0.0006 & 0.5212 & -15\\
         \hline
        2 &  330  & 21,600 & -0.0006 & 0.5212 & -15\\
         \hline
        3 &  330&  14,400 & -0.0006 & 0.5212  & -15\\
        \hline
        4 &  330  &  7,200 & -0.0006 & 0.5212  & -15\\
        \hline
        5 &  500  &  64,696 & 0.00003 & 0.2105 & 10.3\\
        \hline
        6 &  540  &  68,820 & 0.00003 & 0.2144 & 10.3\\
        \hline
        7 &  550 &  100,000 & 0.00003 & 0.2105 & 10.3\\
        \hline
        O\&M & \multicolumn{5}{c|}{0.0218 \$/kWh - For all generators}\\
        \hline
        \end{tabular}
        \vspace{0.1cm}
                \label{Data_Gen_San}
    \end{table}
\end{center}
\vspace{-1cm}
\subsection{New Diesel Generators}
It is assumed that diesel generators may be aggregated in the generation portfolio for load supply and as reserves. Therefore, two types of diesel generators were considered, with their main characteristics being presented in Table \ref{Data_NewDsl}. It was assumed that the minimum load of these generators is also 40\% of their nominal power, i.e., $\psi_{n}=0.4$, as per \cite{[1]} and \cite{[15]}.
\begin{center}
    \begin{table}[t]
     \caption{New diesel generator parameters \cite{[1],[15]}}
        \centering
        \begin{tabular}{|c|c|c|c|c|c|}
        \hline
        \multirow{2}{2em}{Gen.} & Capacity & Lifetime  & a & b & c\\
        & [kW] & [h] & [$l/h/kW^{2}$] & [$l/h/kW$] & [$l/h$]\\ 
        \hline
        1 & 320 &  100,000 & -0.0002 & 0.3287 & 3\\
         \hline
        2 &  520 & 100,000 & -0.00003 & 0.2227 & 10.3\\
         \hline
        Cost & \multicolumn{5}{c|}{727 \$/kW - For all generators}\\
        \hline
        O\&M & \multicolumn{5}{c|}{0.0191 \$/kWh - For all generators}\\
        \hline
        \end{tabular}
        \vspace{0.1cm}
        \label{Data_NewDsl}
    \end{table}
\end{center}
\vspace{-1cm}
\subsection{Solar Panels and Irradiance}
\begin{figure}[t]
    \centering
    \includegraphics[width=1\linewidth]{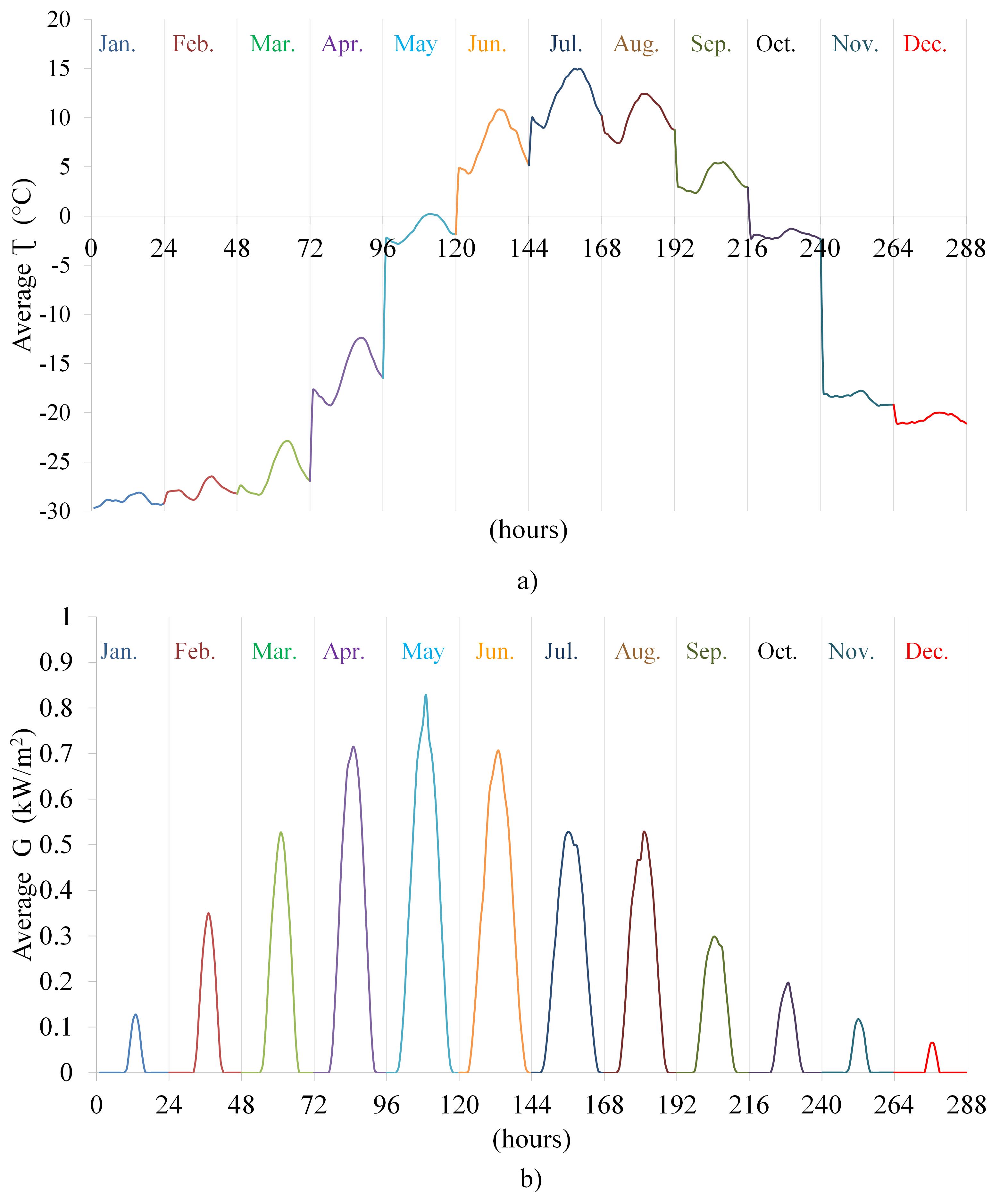}
    \caption{Sanikiluaq's monthly average (a) temperatures $\tau$ and (b) solar irradiance $G$ \cite{[1],[15]}.}
    \label{Temp_GHI}
\end{figure}

The sets of 9.6 kW solar panels are assumed to be connected through an inverter to the \acrshort{MG}. The solar cell temperature $\tau$ and monthly solar irradiance $G$, with their averages, are illustrated in Fig. \ref{Temp_GHI}. The operational parameters and costs associated with the panels are shown in Table \ref{solar_data}.
\begin{center}
\begin{table}[t]
    \caption{Solar panels parameters at Sanikiluaq \cite{[1],[15]} }
    \centering
    \begin{tabular}{ |c|c|c|c|c|c|c| } 
    \hline
    Cost & O\&M & $\alpha$ & $df$ & Lifetime &$\tau_{stc}$ & $G_{stc}$\\
    $[$\$/kW] & [\$/kWh] & [pu/$^\circ C$]  & [\%] & [years] & [$^\circ$C] & [kW/m$^{\rm 2}$] \\
    \hline
     5,082 & 0.0145 & -0.041 & 98 &  20 & 25 & 1\\
    \hline
    \end{tabular}
    \vspace{0.1cm}
    \label{solar_data}
\end{table}
\end{center}
\vspace{-1cm}
\subsection{Wind Turbines and Speed}
\begin{figure}[t]
    \centering
    \includegraphics[width=1\linewidth]{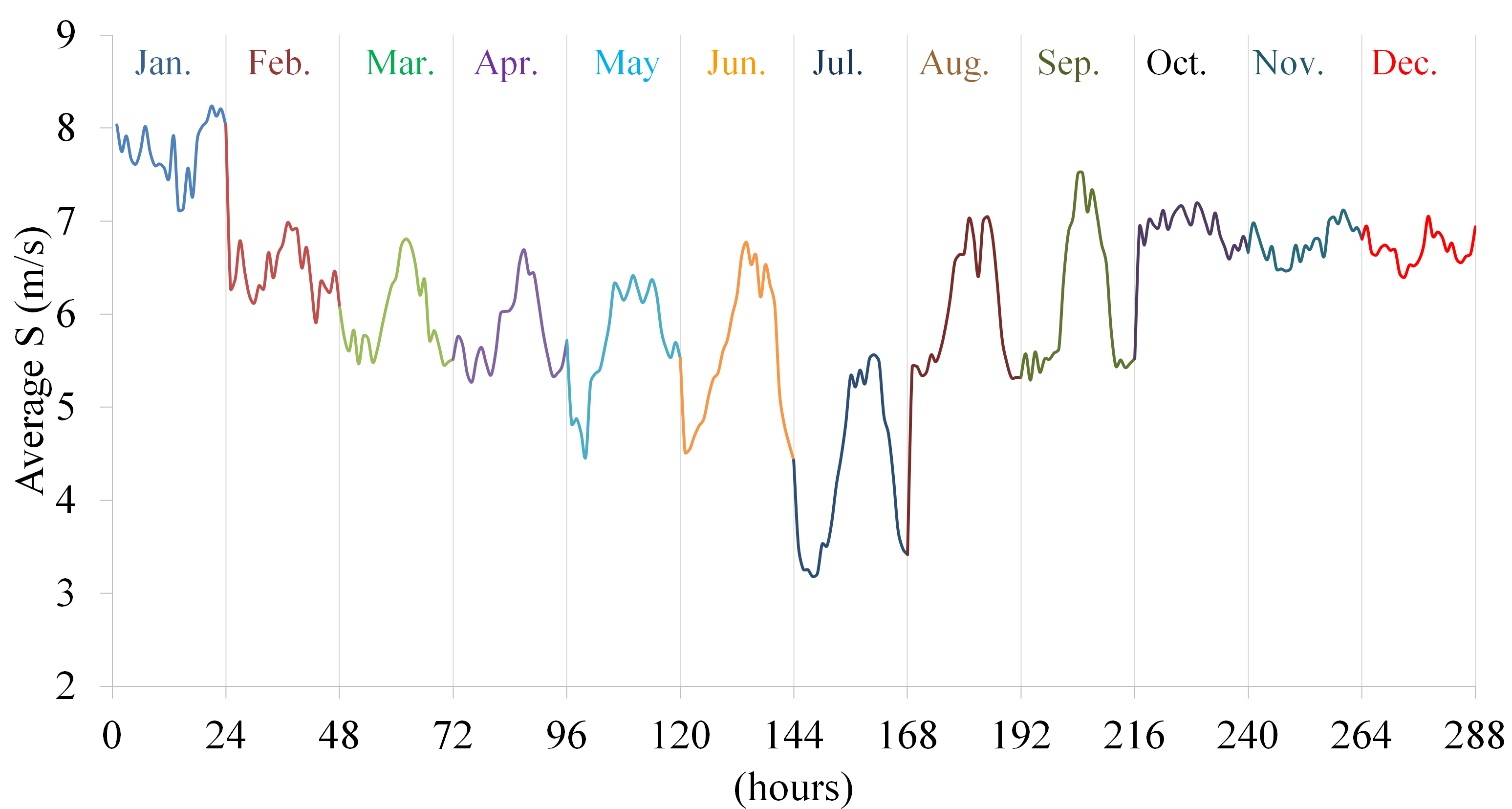}
    \caption{Average wind speed $S$ at 21m hub height \cite{[1],[15]}.}
    \label{fig:Wind_Average}
\end{figure}

Wind generators with 250 kW of nominal capacity were considered with monthly average wind speeds, as shown in Fig. \ref{fig:Wind_Average}. The economical and technical input parameters for the model are presented in Table \ref{wind_data}. The turbine curve $W(\cdot)$ was assumed linear between the cut-in and nominal speed, based on the actual power curves provided in \cite{[15]}, as follows:
\begin{center}
\begin{table}[t]
    \caption{Parameters of wind generators \cite{[1],[15]}}
    \centering
    \begin{tabular}{ |c|c|c|c| } 
    \hline
    Cut-in Speed & Nominal Speed & Cut-out speed & Lifetime\\ 
    $[$m/s] & [m/s] &[m/s] & [years]\\
    \hline
    2.5& 7.5 &25 & 20\\
    \hline
     \multirow{4}{*}{Power Curve} & \multicolumn{3}{l|} {$W(S)=30S-75$ kW \hspace{0.45cm} for $2.5\le S< 5$}\\
     & \multicolumn{3}{l|}{$W(S)=35S-100$ kW \hspace{0.3cm} for $5\le S< 7.5$}\\
     & \multicolumn{3}{l|}{$W(S)=250$ kW \hspace{1.2cm} for $S\le 7.5 <25$}\\
     & \multicolumn{3}{l|}{$W(S)=0$ kW \hspace{1.5cm} Otherwise}\\
    \hline
     Cost & \multicolumn{3}{c|}{7,943 \$/kW}\\
    \hline
    O\&M & \multicolumn{3}{c|}{0.0363 \$/kWh}\\
   \hline
    \end{tabular}
    \vspace{0.1cm}
    \label{wind_data}
\end{table}
\end{center}
\vspace{-1cm}
\subsection{Batteries}
The battery modules in the MG planning model are Li-ion batteries with 100 kWh and 20 kW peak power of charge/discharge, i.e., $t^{ch}=t^{dch}=4h$ for $\delta=0.2$, as per \cite{[1]} and \cite{[15]}. The economical and technical parameters for the implemented battery model are presented in Table \ref{battery_data}.
\begin{center}
\begin{table}[t]
    \caption{Battery parameters \cite{[1],[15]}}
    \centering
    \begin{tabular}{ |c|c|c|c|c|c| } 
    \hline
    Cost & O\&M & $SOC_{0}$ &  $\eta_{Ch}$ & $\eta_{DCh}$\\
    $[$\$/kWh] & [\$/kWh] & [\%] &  [\%] & [\%]\\
    \hline
    1,504 & 0.0069 & 50 & 95 & 95\\
    \hline
    \end{tabular}
    \vspace{0.1cm}
    \label{battery_data}
\end{table}
\end{center}
\vspace{-1cm}
\subsection{Hydrogen System}
To model a hydrogen system, the fuel cells, an electrolyzer, and a hydrogen tank need to be considered. The costs and main characteristics of these elements are presented in Table \ref{Hydrogen_data}.
\vspace{-0.5cm}
\begin{center}
\begin{table}[t]
    \caption{hydrogen system parameters \cite{[17],[14]}}
    \centering
    \begin{tabular}{|c|c|c|c|c|}
    \hline
     Parameter    & Fuel Cells & Electrolizer & Hydrogen Tank \\
     \hline
     Capacity   & 250 kW & 330 kW & 200 kg\\
     \hline
     Cost  & 168,581 \$/u & 1,279,000 \$/u & 249,745 \$/u\\
     \hline
     O\&M  & 2 \$/h & 194 \$/y & 12,400 \$/y\\
     \hline
    Efficiency & $\eta_{FC}=60\%$ & $\eta_{E}=70\%$ & -\\
    \hline
    Lifetime & 50,000 h & 15 y & 25 y\\
    \hline
    \multirow{1}{*}{System} & \multicolumn{3}{c|}{$V=39.4$ kWh, $l_{C}=0.02$ pu} \\
   \multirow{1}{*}{Constants}& \multicolumn{3}{c|}{$\overline{\vartheta}=0.95$ pu,  $\underline{\vartheta}=0.15$ pu } \\

     \hline
    \end{tabular}
     \vspace{0.1cm}
    \label{Hydrogen_data}
\end{table}
\end{center}

\vspace{-0.5cm}
\subsection{Scenarios}
Five scenarios are defined to apply the long term planning model presented in Section II. Note that in order to highlight the contributions of solar generators, each scenario includes one case with solar and one case without solar generators. The cases with solar generation are labeled with A, and the ones without solar are labeled with B. Thus, the main characteristics of these scenarios, considering all possible combinations of \acrshortpl{DER}, are as follows:
\begin{itemize}
    \item  Business-As-Usual (\textit{BAU}) (Base Case): In this case, the only source of generation considered is diesel generation. Other \acrshortpl{DER} are not included here, in order to compare all other scenarios in terms of costs, use of diesel, and GHG reductions. 
    \item 1A ($I$) and 1B ($I-S$): These scenarios include all \acrshortpl{DER}, i.e., diesel ($E,N$), solar ($S$), wind ($W$), batteries ($B$), and hydrogen ($F,Q,\Xi$).
    \item  2A ($I-B$) and 2B ($I-\{B,S\}$): All \acrshortpl{DER} except batteries are considered in this scenario. 
    \item 3A ($I-\{F,Q,\Xi\}$) and 3B ($I-\{F,S,Q,\Xi\}$): All \acrshortpl{DER} except hydrogen storage systems are considered in this scenario.
    \item 4A ($\Pi$) and 4B ($\Pi-S$): In this scenario, only \glspl{RES} and \glspl{ESS} are considered. Diesel generation is considered but exclusively for reserves, to represent a MG supplied primarily by renewable generation. 
\end{itemize}

\subsection{Assumptions and Simulation Criteria}
The \gls{MILP} model, described in Section II, was solved using GAMS \cite{[18]}, with the CPLEX solver. The following are the assumed values for the remaining model parameters \cite{[1],[14],[15]}: 
\begin{itemize}
    \item The discount rate is 8\%.
    \item The planning horizon is 20 years. 
    \item Operation reserves for system adequacy: 50\% for wind ($\rho=0.5$), 25\% for solar ($\gamma=0.25$), and 10\% for load ($\beta=0.1$). 
    \item Load growth is 1.0\%/year. 
    \item \textcolor{black}{It is assumed that the cost of the technology will not be changing throughout the planning horizon, as the balance of the cost associated with the transportation of the equipment and their capital cost may cancel each other for R.C.}
    \item Ramping up/down constraints are not considered, since all diesel generators are able to turn on and off in fractions of an hour.
    \item The cost of diesel is fixed at 2.391 \$/l.
    \item $\mathcal{A}=0.1$ for all diesel generators.
    \item $\mathcal{C}=3000$ cycles of charge and discharge for the batteries.
    \item To control the inclusion of certain \glspl{RES} and \glspl{ESS}, as per the considered scenarios, there must be at least one battery module, 1\% of the annual energy supplied by solar, and/or one hydrogen system module, otherwise the model does not include them due to the cost minimization approach.
    \item The investment in \glspl{RES} is allowed only in the first 5 years \textcolor{black}{to accommodate possible pilot projects}, and new diesel generators are being added from the 3$^{\rm rd}$ to 10$^{\rm th}$ year \cite{[17]}.
    \item For the cases where hydrogen is included in the MG, one full system is included in the first year, leaving the algorithm to decide for additional capacities in the future years. Thus, at least one electrolizer needs to be replaced at year 16, according to their useful lifetime, assuming a zero salvage value.

\end{itemize}

\vspace{-0.2cm}
\section{Results} 

\begin{figure}[t]
    \centering
    \includegraphics[width=\linewidth]{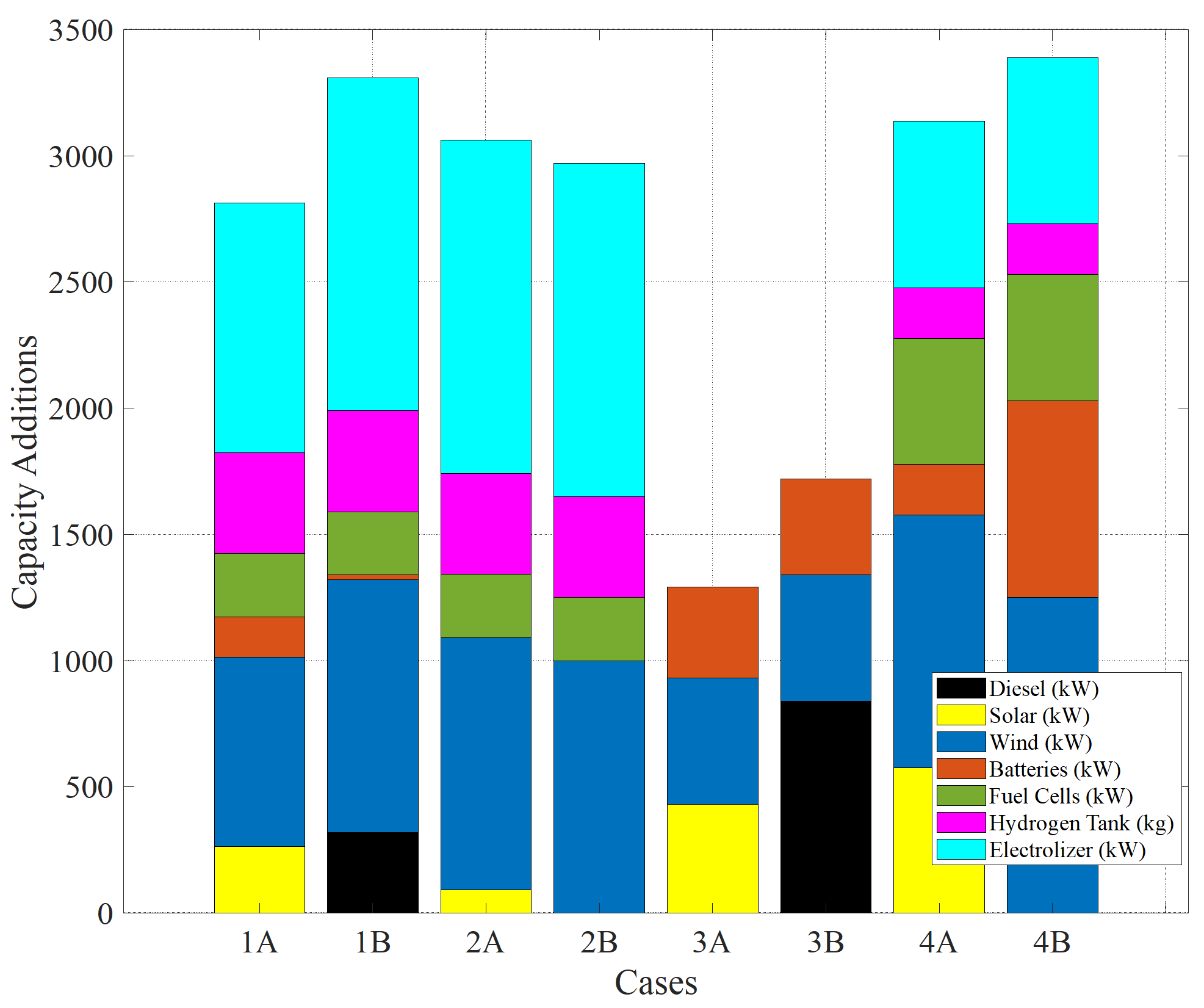}
    \caption{Total capacity additions during the planning horizon.}
    \label{results1}
\end{figure}

\noindent The results of the simulations are shown and discussed in this section. The energy mix resulting from running each scenario can be observed in Fig. \ref{results1}, which illustrates the following: 
\begin{itemize}
    \item Scenarios 1B and 3B, in which solar generation is not considered, recommend investment in diesel generation. Note that larger diesel generation capacities are recommended in Scenario 3B, in which the only source of storage is batteries. Wind generators do not replace solar generation as they have larger capacity, which is not needed to satisfy demand.
    \item In all scenarios, storage capacities of either fuel cells or batteries are used. For example, in the scenarios with only fuel cells (2A and 2B), the energy that has not been served by \glspl{RES} or diesel generators is served by hydrogen storage systems. Also, in Scenarios 3A and 3B, batteries are used, as these are the only available storage capacity.
    \item In Scenarios 1A and 1B, in which the model can choose between investment in hydrogen or batteries, it recommends a portion of both systems.
    \item In Scenarios 4A and 4B, in which diesel generators are not allowed, more investment in storage capacities is recommended to account for the associated uncertainties in the system.
\end{itemize}

\begin{figure}[t]
    \centering
    \includegraphics[width=\linewidth]{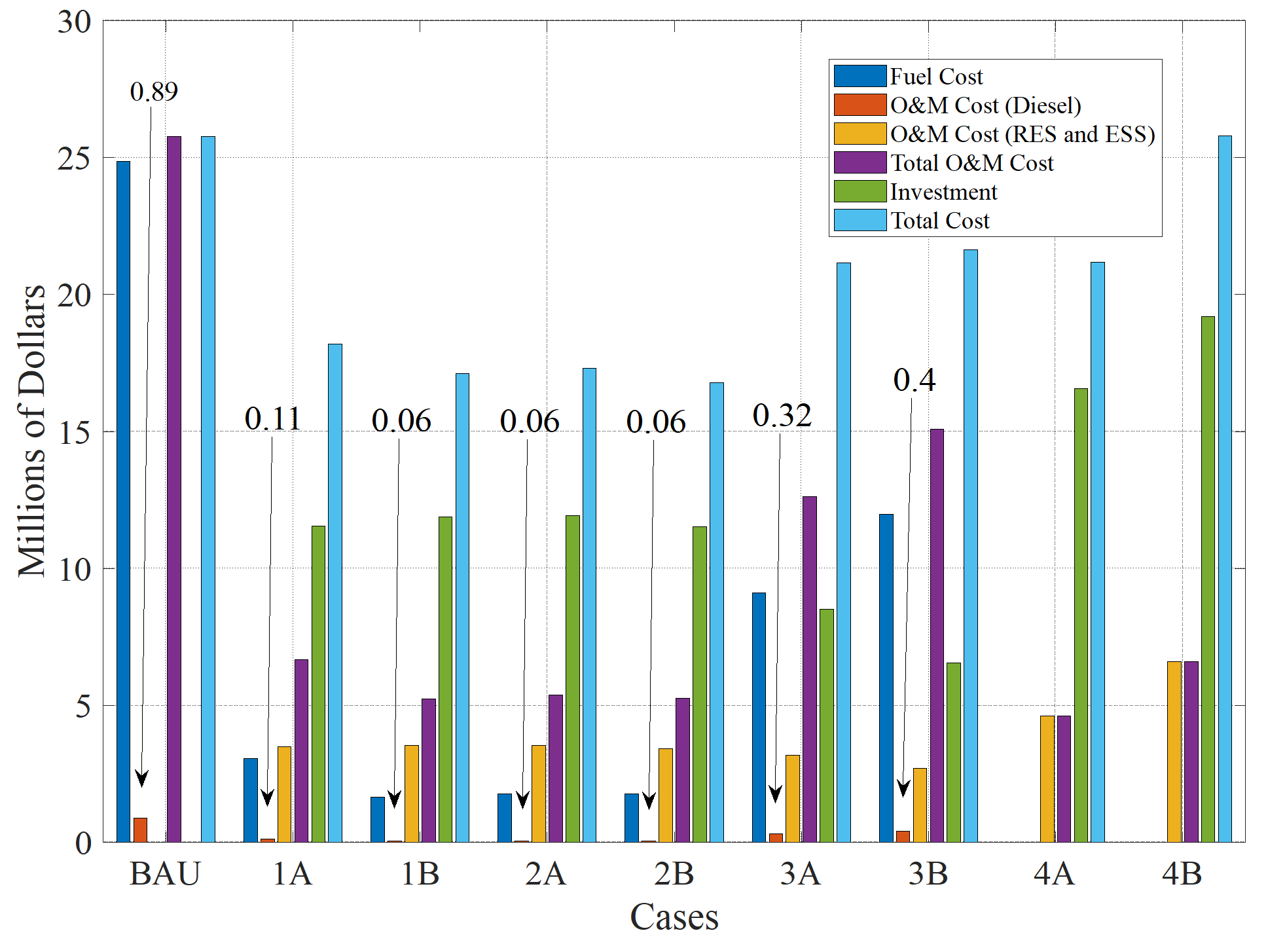}
    \caption{Associated costs of the MG planning for 20 years.}
    \label{results2}
\end{figure}

\begin{figure}[t]
    \centering
    \includegraphics[width=\linewidth]{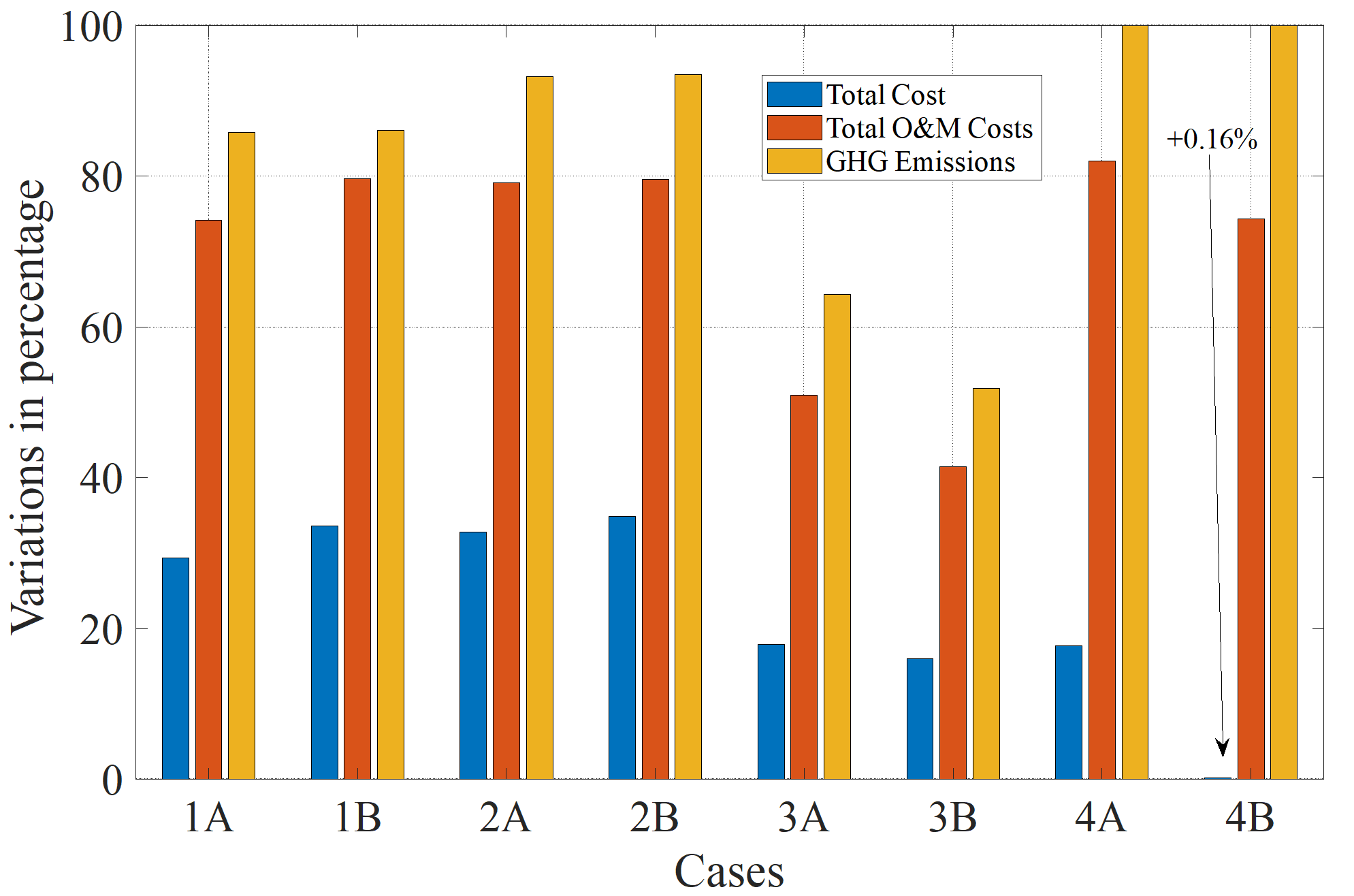}
    \caption{Reduction of cost and emission comparing to the Base Case (BAU)}
    \label{Costs}
\end{figure}

In Figs. \ref{results2} and \ref{Costs} the comparisons among costs and GHG reductions for different scenarios are illustrated. Thus, Fig. \ref{results2} depicts different types of costs associated with each scenario, and Fig. \ref{Costs} illustrates the reductions of total cost, O\&M costs, and GHG reductions in relation to BAU. As observed in Fig. \ref{Costs}, the total O\&M costs decrease from 41\% (3B) to 82\% (4A), and the total costs decrease from 16\% (3B) to 34\% (1B), with respect to BAU. Similarly, the cost of fuel is reduced from 52\% (3B) to 100\% (4A\&4B), with respect to BAU. The most expensive scenario is (4B), in which all renewable resources except solar are recommended, surpassing the total cost of BAU by only 0.16\%, while reducing GHG emissions by 100\%. It can be also observed that the cases with only hydrogen storage systems (2A\&2B) are less expensive than the cases with only batteries as the storage capacity (3A\&3B).

\begin{figure}[t]
    \centering
    \includegraphics[width=1\linewidth]{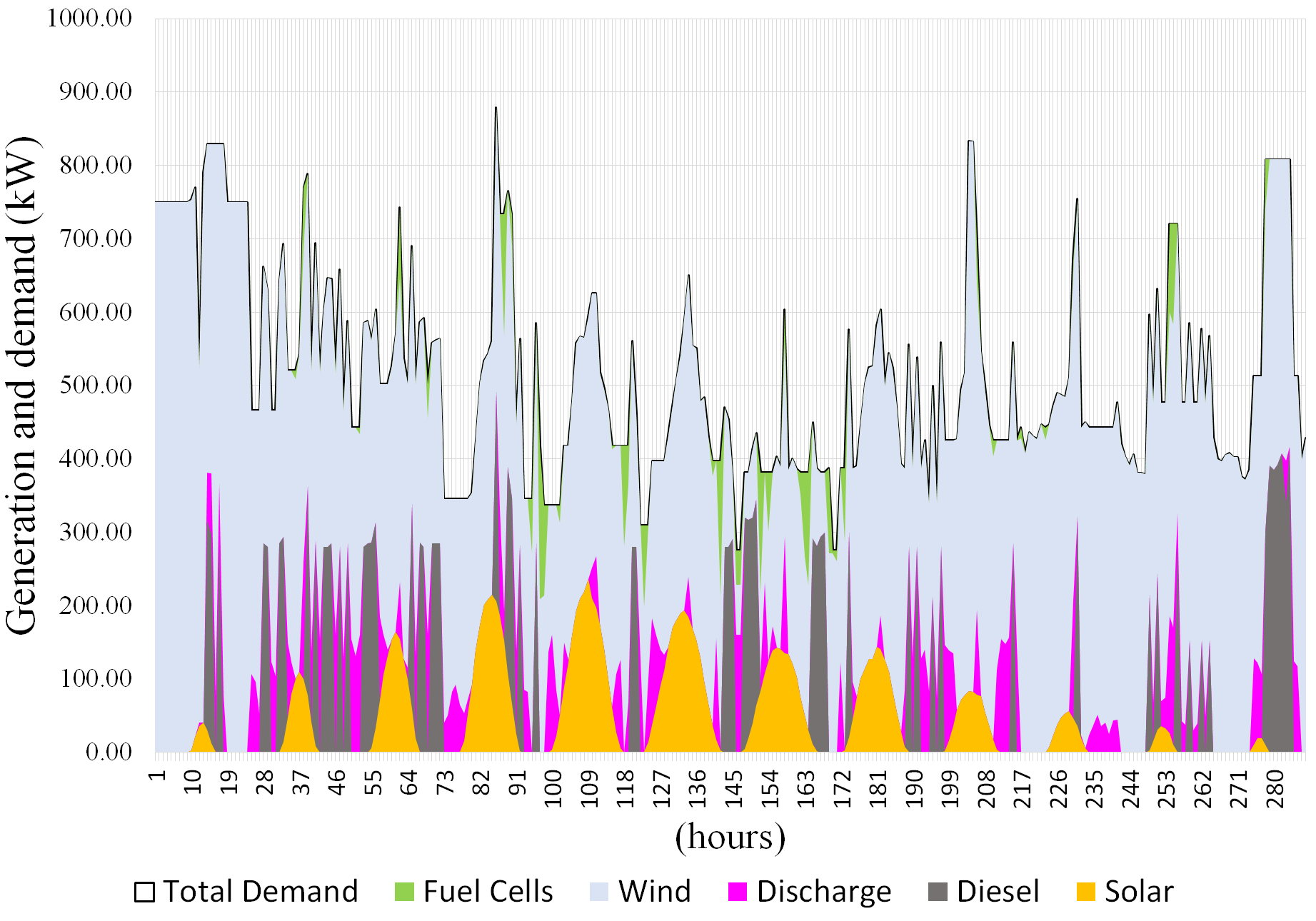}
    \caption{Case 1A operation of the MG for the 10$^{\rm th}$ year.}
    \label{Case1A}
\end{figure}
\begin{figure}[t]
    \centering
    \includegraphics[width=1\linewidth]{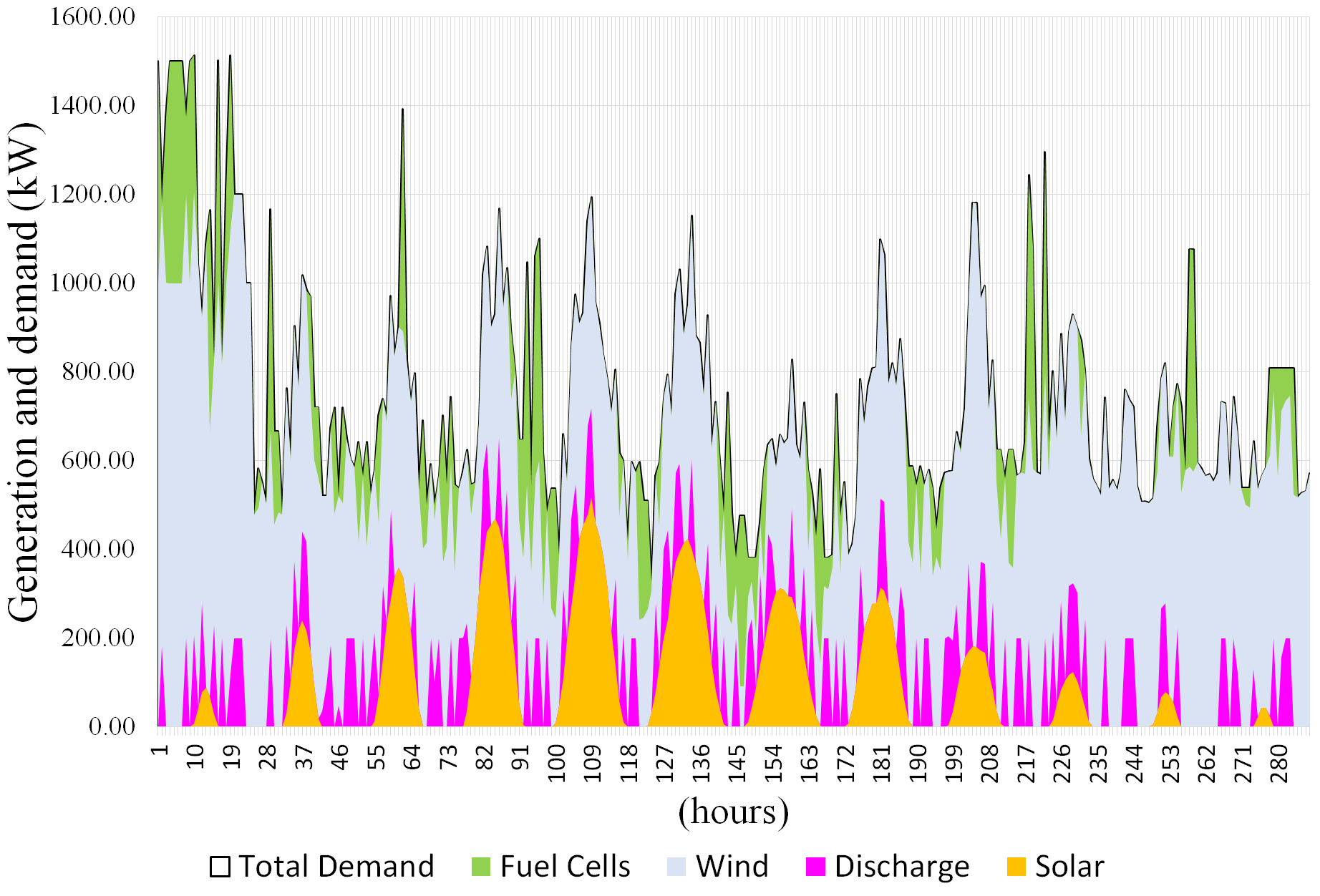}
    \caption{Case 4A  operation of the MG for the 10$^{\rm th}$ year.}
    \label{Case4A}
\end{figure}

Figs. \ref{Case1A} and \ref{Case4A} present the hourly operation of the MG generators and storage systems versus demand, according to Eq. (5), during the 10{$^{\rm th}$} year of its operation for scenarios 1A and 4A, which are chosen as they differ on their type of generation. Note that Scenario 1A allows all \glspl{DER} including diesel, while Scenario 4A allows only \glspl{ESS} and \glspl{RES}. As shown, both batteries and hydrogen systems in combination with other \glspl{DER} are incorporated in the generation mix of the MG to satisfy the hourly demand. In addition, observe that the hydrogen systems can considerably increase the total demand of the MG because of the presence of electrolizers, but the costs can still be reasonable with very low or zero GHG emissions.

\vspace{-0.2cm}
\section{Conclusions}

\noindent It is demonstrated here, based on the proposed \acrshort{MILP} planning model, that the integration of \gls{RES} and \gls{ESS} in the \acrshort{RC} \acrshortpl{MG}' \textcolor{black}{generation portfolio} enhances electric grid flexibility, and promotes decarbonization goals. Thus, as shown in the simulation results, the inclusion of such technologies significantly reduces the use of fossil fuels, resulting in lower emissions (between 51.9\% and 100\%) for the \acrshort{RC} \acrshort{MG} studied. It also not only helps lowering other costs of the energy system, such as fuel storage and its transportation to \acrshortpl{RC}, but also reduces the uncertainties associated with diesel fuel prices. Thus, these results clearly show that wind resources along with solar and storage technologies (batteries and fuel cells) can play a key role in satisfying the electricity demand of \acrshortpl{RC}, while significantly reducing costs and GHG emissions. The model proposed in this paper encourages a structural and economical change, and supports Canada in meeting zero emission targets by introducing \glspl{RES} and \glspl{ESS} in \acrshortpl{RC}, while demonstrating the operational and economic feasibility of such systems.

\bibliography{References.bib}
\bibliographystyle{IEEEtran}

\end{document}